\def\half{\frac{1}{2}}
\def\ph{\phantom{-}} % phantom sign space

\def\eq#1{Eq.~(\ref{#1})}
\def\nl{\hfil\break\noindent}
\documentclass[amsmath,amssymb,superscriptaddress,showkeys,showpacs,nofootinbib]{revtex4}
\usepackage{amsmath}
\usepackage{bm}
\usepackage{graphicx}
\usepackage{rotating}
\usepackage{mathrsfs} %Script characters: {\mathscr D}

\begin{document}
% ------------------------------------------------------------------
%\hspace*{5 in}CUQM - 151 [?]
%\vskip 0.4 in
\hfill HEPHY-PUB 937/14, CUQM 151\vspace*{7ex}
% ------------------------------------------------------------------
\def\ttitle{Schr\"odinger models for solutions of the
Bethe--Salpeter equation in Minkowski space. II. Fermionic
bound-state constituents}\title{\ttitle}
%\markboth{R.~L.~Hall \& W.~Lucha}{\ttitle}

\author{Richard L.~Hall}\email{rhall@mathstat.concordia.ca}
\affiliation{Department of Mathematics and Statistics, Concordia
University, 1455 de Maisonneuve Boulevard West, Montr\'eal,
Qu\'ebec, Canada H3G 1M8}
\author{Wolfgang Lucha}\email{wolfgang.lucha@oeaw.ac.at}
\affiliation{Institute for High Energy Physics, Austrian Academy
of Sciences,\\Nikolsdorfergasse 18, A-1050 Vienna, Austria}

\begin{abstract}In view of the obstacles encountered in any
attempts to solve the Minkowski-space Bethe--Salpeter equation for
bound states of two fermions, we study the possibility to model
the bound-state features, at least at a {\em qualitative\/} level,
by a Schr\"odinger description. Such a nonrelativistic potential
model~can be constructed by applying, to any given Bethe--Salpeter
spectral data, `geometric spectral inversion' in its recently
extended form, which tolerates also {\em singular\/} potentials.
This leads to the adaptation of explicit models that provide an
overview accounting for the Bethe--Salpeter formalism's
complexities.\end{abstract}\maketitle

\section{Introduction: Motivation and Incentive}The Bethe--Salpeter
formalism offers a description of bound states consistent with all
the requirements of relativistic quantum field theory
\cite{BSE_APS,GML,BSE}: bound states with all their properties,
such as their masses or the distribution of the relative momenta
of their constituents (encoded in suitably defined amplitudes
playing the role of a kind of wave functions) arise as solutions
of appropriate equations of motion, called the (homogeneous)
Bethe--Salpeter (BS) equation in the case of bound states composed
of two constituents or the (homogeneous) Faddeev equation for
three-particle systems. In these equations, the underlying
dynamics enter via the full propagators of the constituents and
their interactions kernels --- Green functions which, at least in
principle, may be deduced from the infinite tower of corresponding
Dyson--Schwinger equations. In practice, we are limited to
adequate truncations of this tower and the tight
bounds~of~perturbation~theory.

However, attempts to obtain solutions to the BS equation in
Minkowski space, characterized by the pseudo-Euclidean space-time
metric tensor $g_{\mu\nu}={\rm diag}(+1,-1,-1,-1),$ have to face
serious obstacles in form of singularities induced by propagators
or interaction kernels. The usual remedy is to trust in analytic
continuation and Cauchy's integral~theorem and to formulate the BS
equation in Euclidean space, with metric
$g_{\mu\nu}=\delta_{\mu\nu}.$ However, even if the two approaches
yield the same bound-state masses, predictions based on the
amplitudes describing the constituents' motion differ~drastically.

In view of this, we recently proposed \cite{FICK} to construct, by
spectral inversion of the bound-state energies, approximately
equivalent Schr\"odinger models, and applied \cite{FICK} this
idea, as a kind of feasibility study, to Minkowski-space BS
results for bound states of two \emph{scalar\/} particles. Here,
we extend this analysis to the case of higher relevance for
physics: \emph{fermionic\/} bound-states constituents. Couplings
of fermions to bosons distinguish between bosons of different
Lorentz nature. We use as input BS findings from exchange of a
single scalar, pseudoscalar, or massless vector boson reported in
Refs.~\cite{CK10a,CK10b,CK10c}, and invert the data by our
generalization to singular potentials \cite{FIS} of a previously
formulated inversion technique
\cite{inv1,inv2,inv3,inv4,inv5,inv6}. We may get a first idea of
what to expect by inspecting the BS formalism's nonrelativistic
limit;~see,~e.g.,~Refs.~\cite{Lucha91,Lucha92,Lucha:Oberwoelz,
Lucha:Dubrovnik}.

The outline of this paper is as follows. In Sec.~\ref{Sec:GSI}, we
recall, merely to the extent required for the present
investigation, the principal features of our previously
constructed geometric spectral inversion theory \cite{FIS,FICK}.
In order to prepare the grounds for the applications of this
approach to the energy levels of fermion--(anti-)fermion bound
states resulting from the (Minkowski-space) BS formalism, we
collect, in Sec.~\ref{Sec:CKdata}, the binding-energy predictions
of Refs.~\cite{CK10a,CK10b,CK10c}. In Sec.~\ref{Sec:EP}, we
subject these mass spectra to the inversion procedure. The
insights gained by these efforts finally motivate~us to search, in
Sec.~\ref{mmodels}, for analytic models, based on shifted-Coulomb
or Hulth\'en potentials, rather accurately reproducing~the~data.

\section{Geometric spectral inversion}\label{Sec:GSI}Let us start
by sketching briefly the reasoning of Refs.~\cite{FIS,FICK}
leading to the spectral inversion algorithm and~stating a
uniqueness theorem. The underlying \emph{functional inversion\/}
was first introduced in Ref.~\cite{inv6}. We consider the discrete
spectrum of a Schr\"odinger Hamiltonian operator
\begin{equation}\label{hamiltonian}
H\equiv-\Delta+v\,f(r)\ ,\qquad r\equiv\|\bm{r}\|\
,\qquad\bm{r}\in{\mathbb R}^3\ ,
\end{equation}
where $f(r)$ is an attractive central-potential shape and $v>0$ is
its coupling parameter. Assume that $f(r)$ is monotone
non-decreasing and no more singular than the Coulomb potential
$f(r)=-1/r.$ Then, the operator inequality \cite{GS,RS2}
$H\ge1/(4\,r^2)+v\,f(r)$, along with a simple variational upper
bound to $\langle H\rangle$, enables us to show that a discrete
spectrum exists for sufficiently large coupling $v>v_1>0$. In
particular, the ground-state energy $E$ may be written as a
function $E=F(v)$. We are interested in the problem: can we
reconstruct the potential shape $f(r)$ from the spectral
data,~given by, e.g., the ground-state curve $F(v)$? We call this
type of reconstruction `geometric spectral inversion'.

Discrete spectra of operators bounded from below can be
characterized variationally \cite{RS4}; their ground-state
energy~is
\begin{equation}\label{varchar}
F(v)=\inf_{{{\scriptstyle\psi\in{\cal D}(H)}
\atop{\scriptstyle\|\psi\|=1}}}(\psi,H\psi)\ ,
\end{equation}
where ${\cal D}(H)$ is the domain of $H.$ Define, for the ground
state $\psi$, the kinetic potential $\bar{f}(s)$ associated with
the potential $f(r)$ by a constrained minimization that keeps the
mean kinetic energy $s\equiv\langle-\Delta\rangle$ constant:
\begin{equation}\label{defKP}
\bar{f}(s)\equiv\inf_{{{\scriptstyle\psi\in{\cal D}(H)}\atop
{\scriptstyle\|\psi\|=1}}\atop{\scriptstyle(\psi,-\Delta\psi)=s}}
(\psi,f\psi)\ .
\end{equation}
A final minimization over $s$ allows us to recover the eigenvalue
$F(v)$ of $H$ from $\bar{f}(s)$:
\begin{equation}\label{efroms}
F(v)=\min_{s>0}\!\left[s+v\,\bar{f}(s)\right].
\end{equation}
The spectral function $F(v)$ turns out to be concave (i.e.,
$F''(v)<0$) and can been shown \cite{inv1} to satisfy
\begin{equation}\label{convexities}
F''(v)\,\bar{f}''(s)=-\frac{1}{v^3}<0\ .
\end{equation}
Hence, $F(v)$ and $\bar{f}(s)$ have opposite convexities and,
moreover, are related by a Legendre transformation
$\bar{f}\leftrightarrow F$~\cite{GF}:
\begin{align}
&\bar{f}(s)=F'(v)\ ,\qquad s=F(v)-v\,F'(v)\label{legendre1}\ ,\\
&\frac{1}{v}=-\bar{f}'(s)\ ,\qquad
\frac{F(v)}{v}=\bar{f}(s)-s\,\bar{f}'(s)\label{legendre2}\ .
\end{align}
$F(v)$ is not necessarily monotone, but $\bar{f}(s)$ is monotone
decreasing. By Eq.~(\ref{legendre1}), in place of $s$ also the
coupling, labelled $u$ for this purpose, may be used as
minimization parameter.

A different formulation of this minimization is found if changing
the kinetic-energy parameter from $s$ to $r$ itself, by inverting
the (monotone) function $\bar{f}(s)$ to define the $K$-function
\begin{equation}\label{Kfunction}
K^{[f]}(r)=s=\left(\bar{f}^{-1}\circ f\right)\!(r)\ ,
\end{equation}
which exhibits invariance with respect to scale and shifts (with
constants $A>0$ and $B$):
\begin{equation}\label{Kinvariance}
K^{[Af+B]}(r)=K^{[f]}(r)\ .
\end{equation}
In terms of $K$, Eq.~(\ref{efroms}) becomes
\begin{equation}\label{efromr}
F(v)=\min_{r>0}\!\left[K^{[f]}(r)+v\,f(r)\right].
\end{equation}
Clearly, $K$ still depends on $f,$ but Eq.~(\ref{efromr}) has
$F(v)$ on one side and $f(r)$ on the other. By inversion of this
relation,~we may accomplish $F\rightarrow f.$ To this end, we
construct a sequence of approximate $K$-functions which do
\emph{not\/} depend on $f.$

Now, suppose that a $f(r)$ may be written as smooth transformation
$f(r)=g(h(r))$ of a `basis potential' $h(r)$. Then, the knowledge
of the spectrum of $-\Delta+v\,h(r)$ may by exploited to study the
spectrum of $-\Delta+v\,f(r).$ For definite convexity of the
transformation function $g$, the kinetic-potential formalism
immediately provides energy bounds. This follows from Jensen's
inequality \cite{Jensen}, which we rephrase, for our present goal,
in terms of the kinetic-potential bounds
\begin{equation}\label{kpineq}
g''\ge0\qquad\Longrightarrow\qquad\bar{f}(s)\ge g(\bar{h}(s))\
;\qquad g''\le0\qquad\Longrightarrow\qquad\bar{f}(s)\le
g(\bar{h}(s))\ .
\end{equation}
For these, we write $\bar{f}(s)\approx g(\bar{h}(s)),$ where the
symbol $\approx$ is understood to indicate the appropriate
inequality whenever~$g$ has definite convexity. Expressed in terms
of $K$-functions, the above results read
\begin{equation}\label{kg}
K^{[f]}=\bar{f}^{-1}\circ f\approx(g\circ\bar{h})^{-1}\circ(g\circ
h)=\bar{h}^{-1}\circ h=K^{[h]}\ .
\end{equation}
Thus, $K^{[f]}\approx K^{[h]}$ is the approximation sought; it no
longer depends on $f.$ The corresponding energy bounds~become
\begin{equation}\label{eapprox}
E=F(v)\approx\min_{s>0}\!\left[s+v\,g\!\left(\bar{h}(s)\right)\right]
=\min_{r>0}\!\left[K^{[h]}(r)+v\,f(r)\right].
\end{equation}

For an eigenvalue $E$ of $H$ known as function $E=F(v)$ of the
coupling $v>v_1,$ the kinetic potential $\bar{f}(s)$ is found~by
inverting the Legendre transformation (\ref{legendre1}):
\begin{equation}\label{fbarfromF}
F(v)=\min_{s>0}\!\left[s+v\,\bar{f}(s)\right]\qquad\Longrightarrow\qquad
\bar{f}(s)=\max_{v>v_1}\!\left[\frac{F(v)}{v}-\frac{s}{v}\right].
\end{equation}
Furthermore, we also have to invert the relation (\ref{efromr})
between $F^{[n]}$ and $K^{[n]}$:
\begin{equation}\label{KfromF}
K(r)=\max_{v>v_1}\!\left[F(v)-v\,f(r)\right].
\end{equation}

We implement the inversion procedure by starting from a suitably
chosen seed potential shape, $f^{[0]}(r),$ from which~we generate
a sequence $\{f^{[n]}(r)\}_{n=0}^\infty$ of improving
approximations to the potential. The idea behind this is to arrive
at~a map $g$ such that $g(f^{[n]}(r))$ is close to $f(r)$ in the
sense that the arising eigenvalue is close to our starting
point~$F(v).$ At each stage, the envelope approximation is used.
At stage $n$, the best transformation $g^{[n]}$ is deduced by
using the~current potential approximation $f^{[n]}(r)$ as envelope
basis. Thus, each step in the generation of the
sequence~$\{f^{[n]}(r)\}_{n=0}^\infty$~reads
\begin{equation}
\bar{f}=g^{[n]}\circ\bar{f}^{[n]}\qquad\Longrightarrow\qquad
g^{[n]}=\bar{f}\circ\left(\bar{f}^{[n]}\right)^{-1}
\qquad\Longrightarrow\qquad f^{[n+1]}=g^{[n]}\circ
f^{[n]}=\bar{f}\circ K^{[n]}\ .
\end{equation}
In more detail, the ultimate procedure of our inversion algorithm
may be summarized symbolically in the following~way:\begin{align}
f^{[n]}(r)\qquad\longrightarrow\qquad&F^{[n]}(v)\nonumber\\
\qquad\longrightarrow\qquad&K^{[n]}(r)
=\max_{u>v_1}\!\left[F^{[n]}(u)-u\,f^{[n]}(r)\right]\nonumber\\
\qquad\longrightarrow\qquad&f^{[n+1]}(r)
=\max_{v>v_1}\!\left[\frac{F(v)}{v}-\frac{K^{[n]}(r)}{v}\right].
\label{algor}\end{align}For the step $f^{[n]}(r)\longrightarrow
F^{[n]}(v)$ need to know $E=F^{[n]}(v),$ which we get by solving
$\left(-\Delta+v\,f^{[n]}\right)\psi=E\,\psi$ numerically.

The potential shape $f(r)$ is severely constrained by knowledge of
$F(v).$ Consider a singular potential $f(r)$ of the~form
\begin{equation}\label{formf}
f(r)=\frac{g(r)}{r}\ ,
\end{equation}
where $g(0)<0$, $g'(r)\ge0$, and $g(r)$ is not constant. Examples
of such singular shapes $f(r)$ are Yukawa, $g(r)=-e^{-a\,r},$
Hulth\'en, $g(r)=r/(e^{a\,r}-1),$ and linear-plus-Coulomb,
$g(r)=-a+b\,r^2,$ with $a,b>0.$ For this class of potentials, we
have proved \cite{FIS} the following

\nl{\bf Theorem:} \emph{the potential $f(r)$ in
$H=-\Delta+v\,f(r)$ is uniquely determined by its ground-state
energy~function~$E=F(v).$}

\section{Spectral data from Minkowski-space Bethe--Salpeter
equation for fermionic bound-state constituents}\label{Sec:CKdata}
\subsection{Bethe--Salpeter bound-state energies as input data to
spectral inversion}In their discussion \cite{CK10a,CK10b,CK10c} of
the homogeneous Bethe--Salpeter equation in Minkowski space,
Carbonell and Karmanov consider~bound states of two fermionic
constituents of equal masses $m,$ bound by exchanging between
these constituents a single boson of mass $\mu$ and of either
scalar, or pseudoscalar, or vector Lorentz nature. In
Table~\ref{Tab:CK-fermi}, we reproduce,~from
Refs.~\cite{CK10a,CK10b,CK10c}, the five sets of associated
binding energies, $E,$ presently available in the literature,
computed by numerical solution of the corresponding
Minkowski-space Bethe--Salpeter equation, versus our coupling
parameter, $v,$ and, for~the sake of later ease of reference,
grasp the opportunity to label these data sets in a mnemonic way
(last row of Table~\ref{Tab:CK-fermi}).

\begin{table}[ht]\caption{Couplings $v$ and binding energies $E$
arising from (Minkowski-space) Bethe--Salpeter equations
describing bound~states of fermionic constituents of mass $m=1,$
computed for two-fermion systems bound by exchange of a single
scalar or pseudoscalar boson of mass $\mu=0.15$ or $\mu=0.5$ and
for fermion--antifermion systems bound by a single massless
vector-boson exchange
\cite{CK10a,CK10b,CK10c}.\protect\footnote{Note that inspection of
the nonrelativistic limit reveals that our coupling, $v,$ is
related by $v=g^2/(4\,\pi)$ to the coupling $g$ used
in~Refs.~\cite{CK10a,CK10b,CK10c}.} The numerical values of the
binding energies $E$ have been computed for the choices
$\Lambda=2$ for the vertex form-factor parameter~$\Lambda$ in the
vertex form factor $F(k)$ of Eq.~(\ref{Eq:VFF}) and $L=1.1$ for
the `mass' $L$ in the `discontinuity-smoothing' factor~$\eta(p,P)$
of Eq.~(\ref{Eq:DSF}).}\label{Tab:CK-fermi}
\begin{center}\begin{tabular}{llcllclr}\hline\hline\\[-1.5ex]
\multicolumn{7}{c}{$v$}&\multicolumn{1}{c}{$E$}\\[1.5ex]
\cline{1-7}\\[-1.5ex]\multicolumn{5}{c}{Fermion--Fermion Bound
State}&&\multicolumn{1}{c}{Fermion--Antifermion Bound
State}\\[1.5ex]\cline{1-5}\cline{7-7}\\[-1.5ex]
\multicolumn{2}{c}{Scalar-Boson Exchange}&&
\multicolumn{2}{c}{Pseudoscalar-Boson Exchange}&&
\multicolumn{1}{c}{Vector-Boson Exchange}\\[1.5ex]
\cline{1-2}\cline{4-5}\cline{7-7}\\[-1.5ex]
\multicolumn{1}{l}{$\mu=0.15\quad$}&\multicolumn{1}{l}{$\mu=0.50$}&&
\multicolumn{1}{l}{$\mu=0.15\quad$}&\multicolumn{1}{l}{$\mu=0.50$}&&
\multicolumn{1}{l}{$\mu=0$}\\[1.5ex]\hline\\[-1.5ex]
$0.6217$&$2.008$&&$17.89$&$33.61$&&$0.2598$&$-0.01$\\
$0.7998$&$2.347$&&$18.53$&$34.23$&&$0.3907$&$-0.02$\\
$0.9510$&$2.627$&&$18.80$&$34.68$&&$0.4984$&$-0.03$\\
$1.089$&$2.880$&&$19.35$&$35.05$&&$0.5934$&$-0.04$\\
$1.222$&$3.119$&&$19.66$&$35.36$&&$0.6802$&$-0.05$\\
$1.840$&$4.203$&&$20.86$&$36.60$&&$1.046$&$-0.10$\\
$3.049$&$6.227$&&$22.51$&$38.25$&&$1.626$&$-0.20$\\
$4.313$&$8.260$&&$23.76$&$39.58$&&$2.109$&$-0.30$\\
$5.656$&$10.40$&&$24.81$&$41.00$&&$2.534$&$-0.40$\\
$6.919$&$12.53$&&$25.71$&$41.85$&&$2.914$&$-0.50$\\
[1.5ex]\hline\\[-1.5ex] data $\mbox{S}_1$&data $\mbox{S}_2$&&data
$\mbox{P}_1$&data $\mbox{P}_2$&&data $\mbox{V}$&\\
\\[-1.5ex]\hline\hline\end{tabular}\end{center}\end{table}

\subsection{Additional complications in the case of fermionic
bound-state constituents}In their numerical studies, Carbonell and
Karmanov find, at least for both scalar- and vector-boson
exchanges,~that beyond some \emph{critical\/} value $g_{\rm c}$ of
the respective \emph{coupling constant\/} $g,$ namely, $g_{\rm
c}=2\,\pi$ in the scalar-boson case and~$g_{\rm c}=\pi$ in the
vector-boson case, the resulting energy spectrum is
\emph{unbounded from below\/}. This --- strange but for good
reasons beyond doubt highly unwanted --- feature is cured by
introduction of a vertex form factor $F(k),$ $k\equiv p-q,$ of the
form\begin{equation}F(k)=\frac{\mu^2-\Lambda^2}{k^2-\Lambda^2+{\rm
i}\,\varepsilon}\ ,\qquad\lim_{\Lambda\to\infty}\!F(k)=1\
,\label{Eq:VFF}\end{equation}and by regularization of the relevant
interaction vertices by replacing the corresponding coupling
constant $g$ by $g\,F(k).$ Moreover, the interaction kernels
exhibit a discontinuous behaviour as functions of the respective
integration variables. These discontinuities are smoothed by
multiplying the entire integral equation, that is, both LHS and
RHS, by a factor $\eta(p,P)$ which is reminiscent of the product
of the free propagators of two spinless particles carrying~momenta
$p_1$ and~$p_2$:
\begin{equation}\eta(p,P)=\frac{m^2-L^2}{(\half P+p)^2-L^2+{\rm
i}\,\varepsilon}\,\frac{m^2-L^2}{(\half P-p)^2-L^2+{\rm
i}\,\varepsilon}\ ,\qquad\lim_{L\to\infty}\!\eta(p,P)=1\ ,\qquad
L>m\ .\label{Eq:DSF}\end{equation}Carbonell and Karmanov claim
that the latter modification does {\em not\/} change the resulting
solutions of the BS equation.

\section{Construction of effective potentials by geometric spectral
inversion}\label{Sec:EP}
\begin{figure}[ht]\centering\includegraphics[scale=.5447]{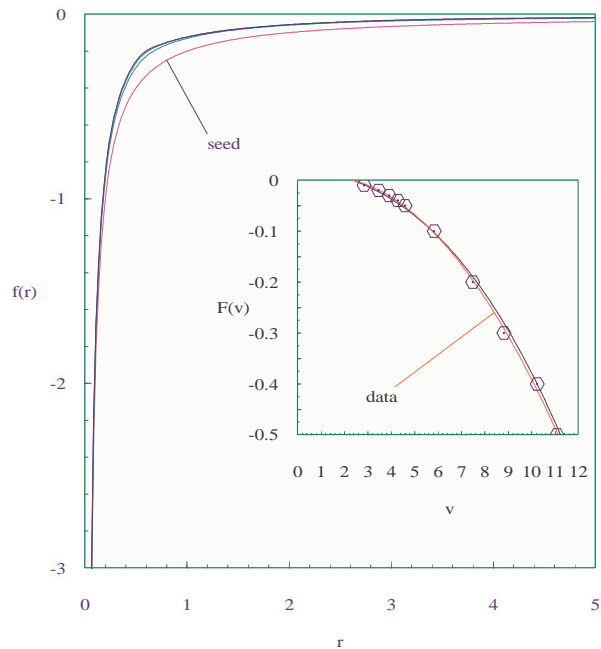}
\caption{Geometric inversion of the data set $\mbox{P}_2$ of
fermion--antifermion binding energies in Table~\ref{Tab:CK-fermi}.
We depict, for the potential $f(r),$ a sequence of three
iterations of inversion, $f^{[k]}(r),$ $k=0,1,2,3,$ where the
seed, $f^{[0]}(r)=-0.2/r,$ is the lowest (red) curve, as well as,
in the insert graph, both the interpolation curve $F(u)$ of the
binding-energy input (hexagons) and the corresponding eigenvalue
curve of the Hamiltonian $H=-\Delta+u\,f^{[3]}(r).$ Since $E=0$
for $v_0\approx30.8,$ the sequence inverts $F(u),$~where~$u\equiv
v-v_0.$ (Note that $F(u)$ is positive in the interval
$u\in(0,u_c),$ $u_c=4\,b/a^2=2.3726,$ rising to a maximum of
$0.0072$ at the interval~center.)}\label{figd54}\end{figure}

We now use our inversion theory \cite{FIS}
to find the potential shape $f(r)$ in the Schr\"odinger
Hamiltonian $H,$ Eq.~(\ref{hamiltonian}),~that, for any spectral
data set in Table~\ref{Tab:CK-fermi}, generates those binding
energies $E$ for the given values of the coupling parameter~$v$.
Let us start our inversion algorithm, \eq{algor}, from a pure
Coulomb seed potential $f^{[0]}(r)=-\alpha/r,$ with
constant~$\alpha>0$. Since there are merely ten points for each
data set, it is necessary first to represent each energy curve
$F(v)$ by a~smooth interpolating function. Each data set
determines a cutoff value $v_0$ of the coupling, defined by
$F(v_0)=0.$ In cases such as data $\mbox{P}_2$ of Table
\ref{Tab:CK-fermi}, where $v_0\approx30.8$ is large, we found that
the inversion algorithm converges very slowly. It became clear
that it was much more effective to shift the abscissae of the data
by using the new variable $u\equiv v-v_0$, in effect inverting
$\mathcal{F}(u)\equiv F(v_0+u)$ instead of $F(v)=F(v_0+u).$ Thus,
using the data set $\mbox{P}_2$ of Table \ref{Tab:CK-fermi}, we
find the potential shape whose graph is plotted in
Fig.~\ref{figd54}: an abuse of notation allows us to label the
energy ordinate~generically as~$F(u).$

\vspace{-5.315ex}

\section{Minimal Schr\"odinger models}\label{mmodels}
\subsection{Coulomb model}\label{Sec:cmodel}Encouraged by the
effectiveness and rapid convergence of the inversion of
Sec.~\ref{Sec:EP}, we adopt tentatively the following simplified
method. Consider the Schr\"odinger Hamiltonian $H$ for the
relative energy of two particles of common mass~$m$,
\begin{equation}\label{Eq:cmodel}
H=-\frac{1}{m}\,\Delta+(v-v_0)\left(-\frac{a}{r}+b\right),
\end{equation}
where the three adjustable parameters $\{a,b,v_0\}$ are the
Coulomb weight $a,$ a potential shift $b,$ and the critical
coupling for vanishing energy, $v_0.$ All corresponding exact
eigenenergies $E_{n\ell}(v)$ are immediately given by the
elementary~formula
\begin{equation}\label{eformula}
E_{n\ell}(v)=b\,(v-v_0)-\frac{m\,a^2\,(v-v_0)^2}{4\,(1+n+\ell)^2}\
,\qquad n=0,1,2,\dots\ ,\qquad\ell=0,1,2,\dots\ .
\end{equation}
For the case at hand, we have $m=1$ for the mass, we set $u\equiv
v-v_0$, and we consider the ground state (identified~by~the
quantum numbers $n=\ell=0$) for which we obtain the energy formula
$E=F(u)$ and its exact inversion $f(r)$ as follows:
\begin{equation}\label{egformula}
F(u)=-\frac{a^2\,u^2}{4}+b\,u\qquad\longrightarrow\qquad
f(r)=-\frac{a}{r}+b\ .
\end{equation}
The idea is that we fit the energy formula to the given energy
data by finding the best parameter triple $\{a,b,v_0\}$:~this, in
turn, specifies the associated potential shape, $f(r)$. The fast
convergence of the inversion algorithm that we exhibited
numerically in Fig.~\ref{figd54} is, in fact, realized
analytically in just a single step, as we now show in the
following~subsection.

\subsection{Exact inversion in one step}We prove, for the inversion
algorithm of Sec.~\ref{Sec:GSI}, that if the seed is of Coulombic
shape, $f^{[0]}(r)=-\alpha/r,$ then~the~first iteration of the
algorithm yields $f^{[1]}(r)=-a/r+b.$ To show this, we simply
apply the inversion algorithm, as follows:\begin{align*}
H^{[0]}=-\Delta-\frac{\alpha\,u}{r}\qquad\Longrightarrow\qquad
&F^{[0]}(u)=-\frac{(\alpha\,u)^2}{4}\\\qquad\longrightarrow\qquad
&K^{[0]}(r)=\max_{u>0}\left[F^{[0]}(u)-u\,f^{[0]}(r)\right]=
\frac{1}{r^2}\\\qquad\longrightarrow\qquad&f^{[1]}(r)
=\max_{u>0}\left[\frac{F(u)}{u}-\frac{K^{[0]}(r)}{u}\right]
\\\qquad\Longrightarrow\qquad&f^{[1]}(r)
=\max_{u>0}\left(-\frac{a^2\,u}{4}-\frac{1}{r^2\,u}+b\right)
=-\frac{a}{r}+b\ .\end{align*}Thus, the inversion $F\rightarrow f$
is achieved in one step. This finding itself is perhaps neither
profound nor surprising,~but~it illuminates our observation, made
in the present context, that geometric spectral inversion is very
efficient when there exists a critical coupling $v_0\ne0$ and the
energy is considered to be a function of $(v-v_0).$ We note in
passing that~the second algorithmic step yields the $K$-function
$K^{[0]}(r)=1/r^2$, which does not depend on the parameter
$\alpha$, even though the seed, $f^{[0]}(r)=-\alpha/r,$ does. This
is consistent with the general invariance of $K$-functions, noted
in \eq{Kinvariance} above.

\subsection{Application of the Coulomb model to Bethe--Salpeter
binding energies}
\begin{figure}[ht]\centering\includegraphics[scale=.5447]{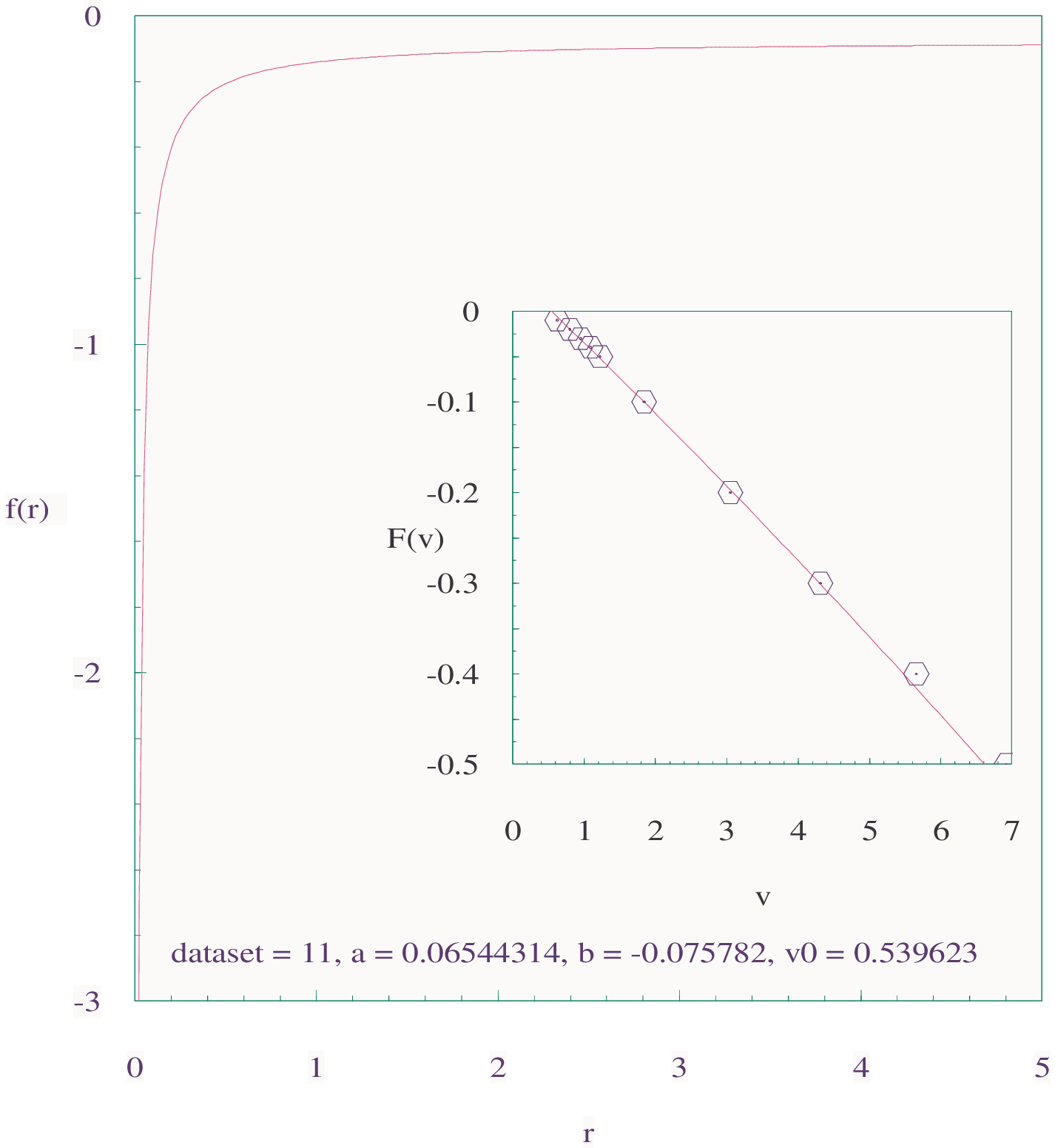}
\caption{Binding-energy data $\mbox{S}_1$ from
Table~\ref{Tab:CK-fermi} (interpolated hexagons) and corresponding
potential shape given by the model~(\ref{egformula}).}
\label{figd11}\end{figure}

\begin{figure}[ht]\centering\includegraphics[scale=.5447]{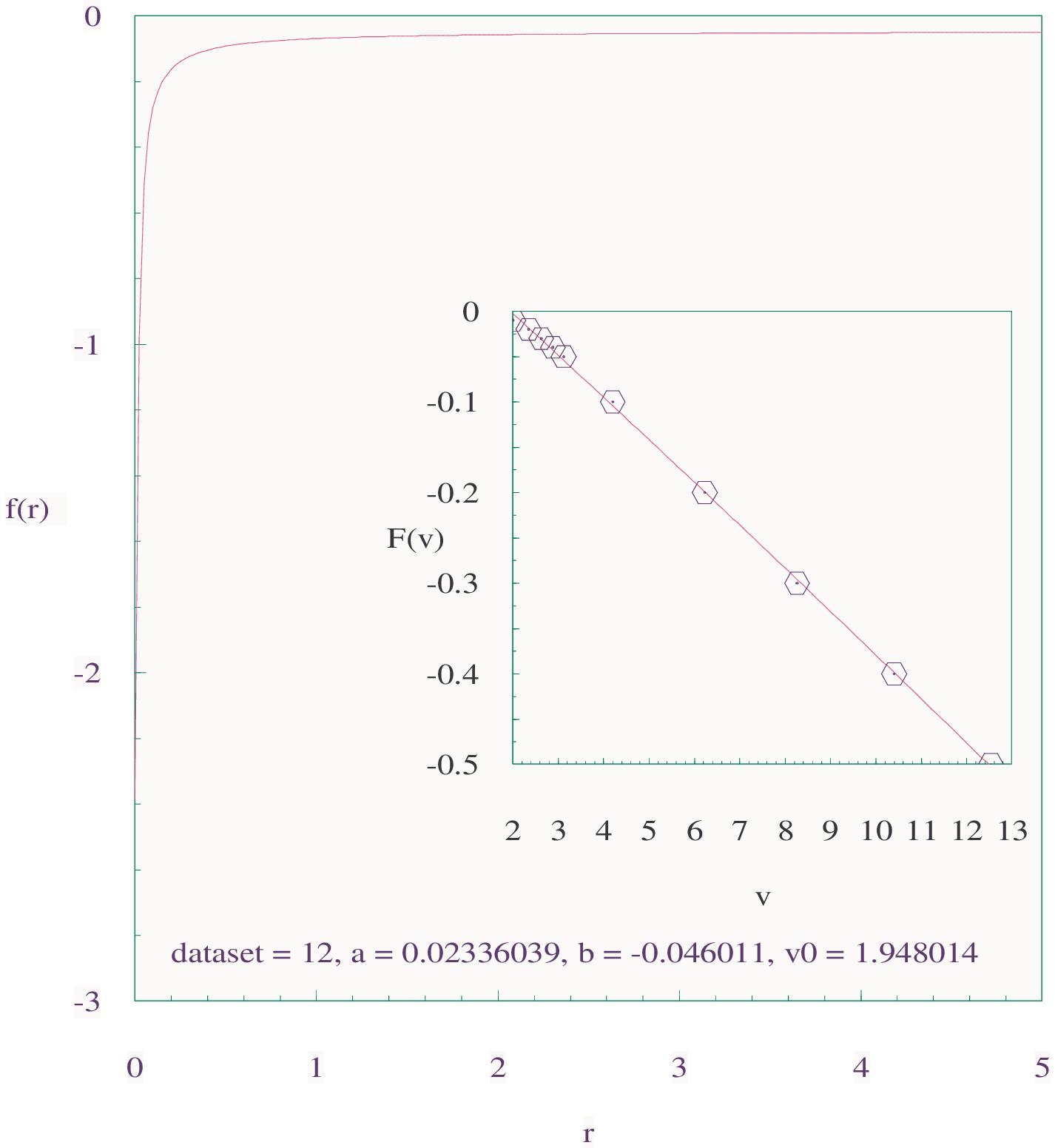}
\caption{Binding-energy data $\mbox{S}_2$ from
Table~\ref{Tab:CK-fermi} (interpolated hexagons) and corresponding
potential shape given by the model~(\ref{egformula}).}
\label{figd12}\end{figure}

\begin{figure}[ht]\centering\includegraphics[scale=.5447]{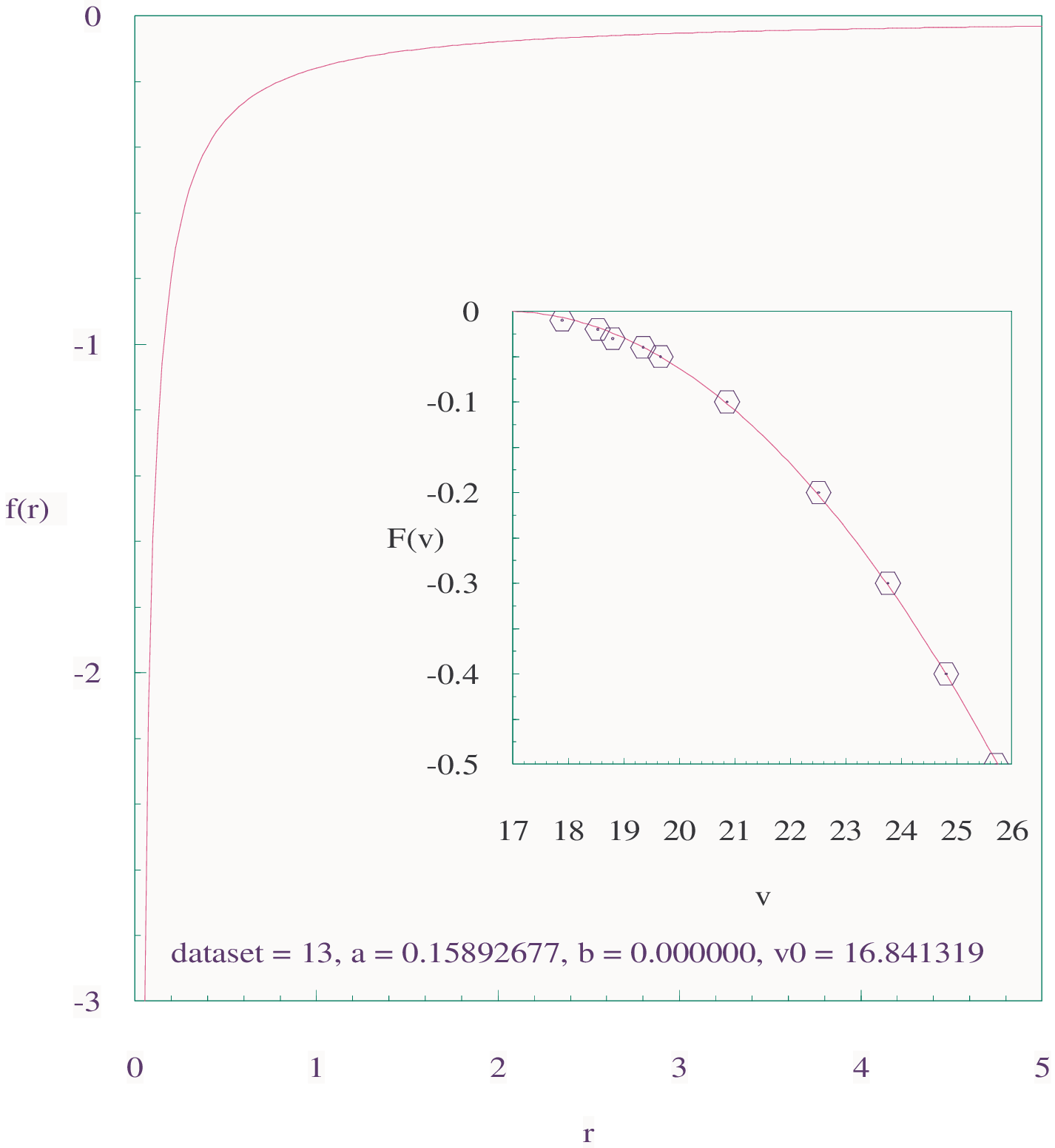}
\caption{Binding-energy data $\mbox{P}_1$ from
Table~\ref{Tab:CK-fermi} (interpolated hexagons) and corresponding
potential shape given by the model~(\ref{egformula}).}
\label{figd13}\end{figure}

\begin{figure}[ht]\centering\includegraphics[scale=.5447]{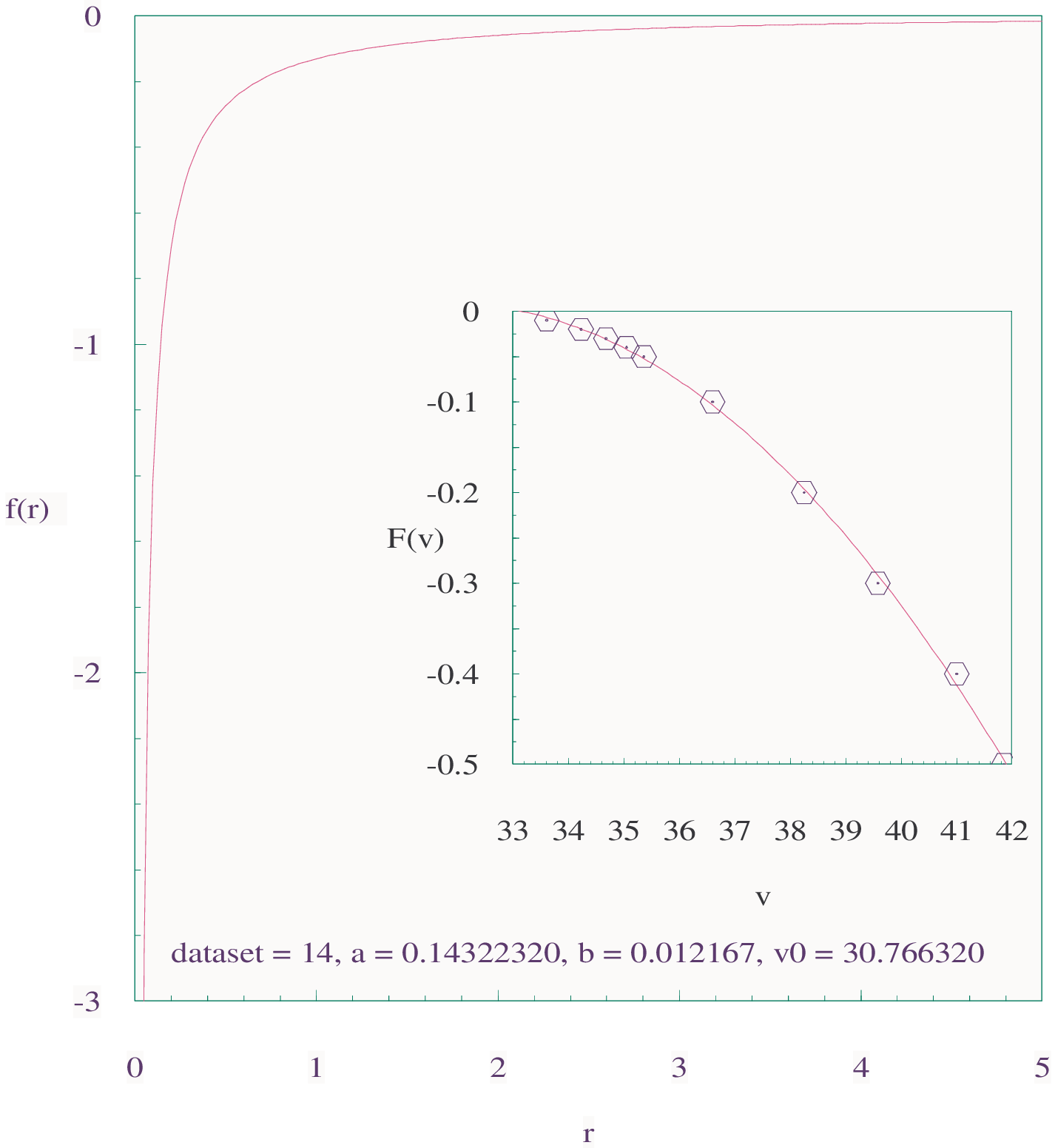}
\caption{Binding-energy data $\mbox{P}_2$ from
Table~\ref{Tab:CK-fermi} (interpolated hexagons) and corresponding
potential shape given by the model~(\ref{egformula}).}
\label{figd14}\end{figure}

\begin{figure}[ht]\centering\includegraphics[scale=.5447]{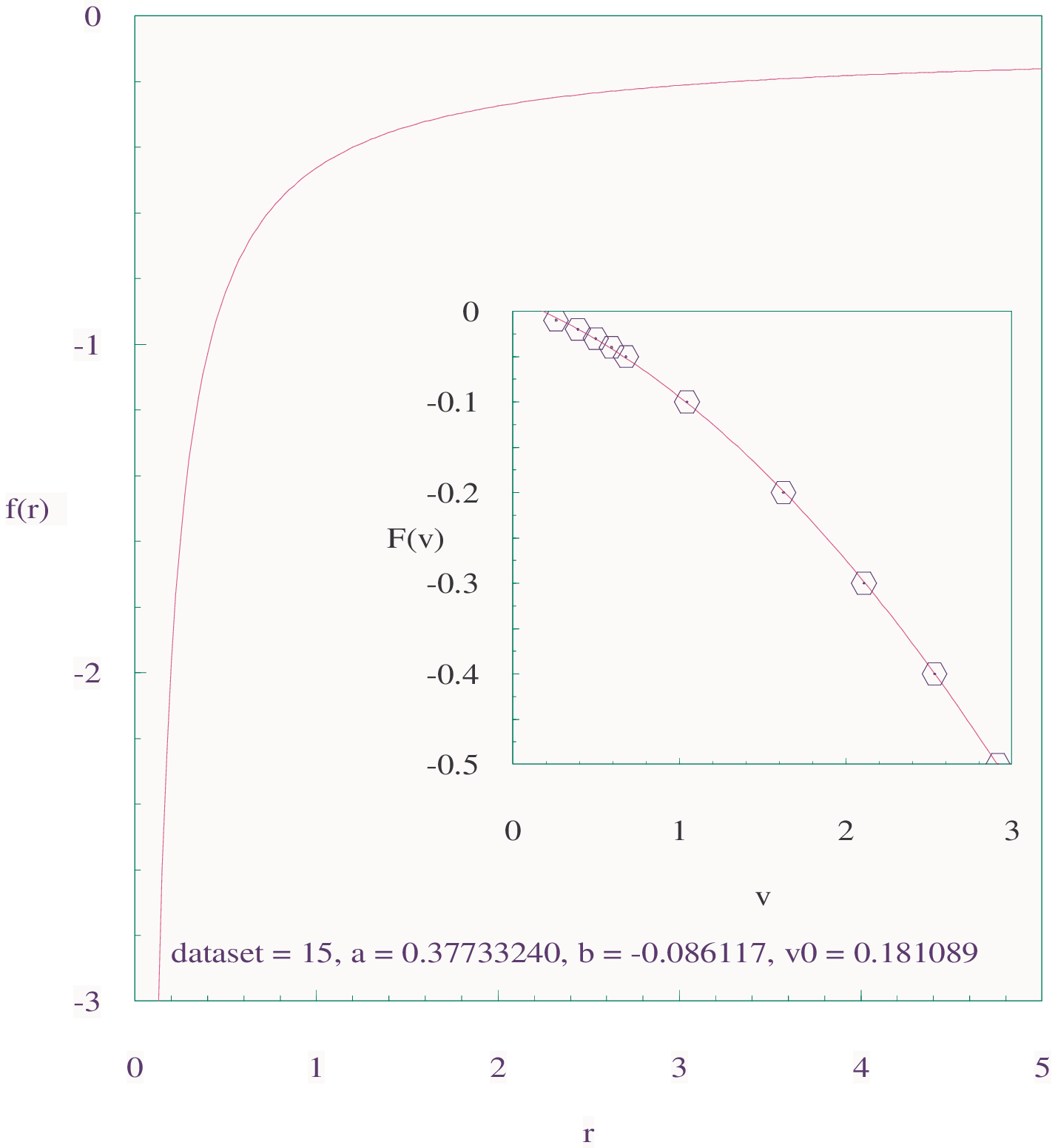}
\caption{Binding-energy data $\mbox{V}$ from
Table~\ref{Tab:CK-fermi} (interpolated hexagons) and corresponding
potential shape given by the model~(\ref{egformula}).}
\label{figd15}\end{figure}

We apply the model (\ref{egformula}) to each of
the data sets in Table~\ref{Tab:CK-fermi}. The fitted values of
the parameters $\{a,b,v_0\}$ are~collected in
Table~\ref{Tab:params} whereas the graphical results for the
potential shapes $f(r)$ and the corresponding $v$-form energy
curves~$F(v)$ are depicted in Figs.~\ref{figd11} through
\ref{figd15}; the interpolated hexagons represent the original
discrete spectral data from Table~\ref{Tab:CK-fermi}.

\begin{table}[h]\caption{Values found for the parameters
$\{a,b,v_0\}$ of the shifted Coulomb potential for each of the
five data sets in Table~\ref{Tab:CK-fermi}.}\label{Tab:params}
\begin{center}\begin{tabular}{cllr}\hline\hline\\[-1.5ex]
Data set&\multicolumn{1}{c}{$a$}&\multicolumn{1}{c}{$b$}&
\multicolumn{1}{c}{$v_0$}\\[1.5ex]\hline\\[-1.5ex]
$\mbox{S}_1$&$0.0654431$&$-0.075782$&$0.539623$\\
$\mbox{S}_2$&$0.0233604$&$-0.046011$ & $1.948014$\\
$\mbox{P}_1$&$0.1589268$&$\ph 0.0$&$16.841319$\\
$\mbox{P}_2$&$0.1432232$&$\ph 0.012167$&$30.76632$\\
$\mbox{V}$&$0.3773324$&$-0.086117$&$0.181089$\\[1.5ex]
\hline\hline\end{tabular}\end{center}\end{table}

\subsection{Equivalent Hulth\'en model}\label{hmodel}As is rather
well known \cite{flug,hallhulthen}, for the (nonrelativistic)
Schr\"odinger Hamiltonian operator with Hulth\'en potential,
\begin{equation}\label{srohul}
H=-\Delta+v\left(-\frac{\alpha}{\exp(\beta\,r)-1}\right),
\qquad\alpha>0\ ,\qquad\beta>0\ ,
\end{equation}
the energy eigenvalues of bound states with vanishing orbital
angular momentum ($\ell=0,$ $s$-states) can be given~exactly:
\begin{equation}E_n=-\frac{\left[v\,\alpha-\beta^2\,(n+1)^2\right]^2}
{[2\,\beta\,(n+1)]^2}\ ,\qquad v\,\alpha\ge\beta^2\,(n+1)^2\
,\qquad n=0,1,2,\dots\ .\end{equation}If we consider the ground
state ($n=0$) and set $\alpha=v_0\,a^2$ and $\beta=v_0\,a,$ then
we arrive at the operator--eigenvalue pair\begin{equation}
-\Delta+v\left(-\frac{v_0\,a^2}{\exp(v_0\,a\,r)-1}\right)
+b\,(v-v_0)\qquad\longrightarrow\qquad
E=-\frac{a^2\,(v-v_0)^2}{4}+b\,(v-v_0)\
.\label{hopepair}\end{equation}We note that Eqs.~(\ref{hopepair})
and (\ref{egformula}) describe the same spectral curve. There is,
however, a qualitative formal difference: in the Coulomb model
(\ref{egformula}), the potential shape $f(r)$ is simply multiplied
by the coupling $(v-v_0)$, whereas, in~the~Hulth\'en Hamiltonian
(\ref{hopepair}), the operator term $b\,(v-v_0)$ contributes the
part $b\,v$ to the potential and simultaneously an additional part
$b\,v_0$ is subtracted from the energy operator at the end. Thus,
although the Hulth\'en model involves a pair potential that is
similar to the Yukawa potential, its use in the context of the
present model structure --- where $(v-v_0)\,f(r)$~so effectively
leads to $F(v)$ --- does not remain our first choice. Of course,
when $b = 0$, this difference between the models is removed, and
on (understandable) grounds of familiarity in the use of coupling,
one might prefer the Hulth\'en option. In any case, both models
are always available and for the ground-state energy they may be
considered to be equivalent.

\section{Conclusion}Numerical solutions \cite{CK10a,CK10b,CK10c} to
the Bethe--Salpeter equation describing bound states of two
fermions yield data for the binding energy $E=F(v)$ as a function
of the coupling parameter $v>0.$ The form of the data suggests
that they may be generated (approximately) by a suitable
nonrelativistic model. Meanwhile, we have at our disposal a
geometric spectral inversion theory
\cite{inv1,inv2,inv3,inv4,inv5,inv6,FIS,FICK} which, if $E=F(v)$
is the lowest eigenvalue of the Schr\"odinger
Hamiltonian~$H\equiv-\Delta+v\,f(r),$ reconstructs from the given
spectral curve $F(v)$ the underlying potential shape $f(r).$ By
first analyzing the fermion data expressed in this manner, as we
had done earlier \cite{FICK} for interacting bosons, we eventually
made an elementary~discovery: namely, when there exists a nonzero
critical value $v_0$ of the coupling $v$ and the model Hamiltonian
is written in the form $H=-\Delta+(v-v_0)\,f(r),$ then the
spectral data $F(v)$ found for the Bethe--Salpeter two-fermion
problem are accurately represented as the eigenvalues of that
Hamiltonian $H$ for which the potential shape has, for $a>0,$ the
elementary form $f(r)=-a/r+b.$ As more eigenvalue data become
available, we expect to be able to translate the essential
features of such relativistic two-particle problems into values
for the parameters $\{a,b,v_0\}$ of this minimal~nonrelativistic
model.

\section*{Acknowledgments}One of us (RLH) gratefully acknowledges
both partial financial support of this work under Grant No.\
GP3438 from the Natural Sciences and Engineering Research Council
of Canada and the hospitality of the Institute for High Energy
Physics of the Austrian Academy of Sciences, Vienna, where part of
this work was done.

\end{document}